\documentclass{article}
\usepackage{spconf,amsmath,epsfig}
\usepackage{url}
\usepackage{graphicx}
\usepackage{subfigure}
\usepackage{booktabs}
\usepackage{bbding}
\usepackage{caption}
\usepackage{amsfonts}
\usepackage{array,multirow}

\let\OLDthebibliography\thebibliography
\renewcommand\thebibliography[1]{
  \OLDthebibliography{#1}
  \setlength{\parskip}{0pt}
  \setlength{\itemsep}{0pt plus 0.3ex}
}

\begin{document}\sloppy

\def\x{{\mathbf x}}
\def\L{{\cal L}}

\title{Selective Residual M-Net for real image denoising}
%

\name{Chi-Mao Fan$^{\star}$ \qquad Tsung-Jung Liu$^{\star}$ \qquad Kuan-Hsien Liu$^{\dagger}$}
\address{
\ninept $^{\star}$Department of Electrical Engineering and Graduate Institute of Communication Engineering, National Chung Hsing University, Taiwan
\\
\ninept $^{\dagger}$Department of Computer Science and Information Engineering, National Taichung University of Science and Technology, Taiwan
}

\maketitle

\begin{abstract}
Image restoration is a low-level vision task which is to restore degraded images to noise-free images. With the success of deep neural networks, the convolutional neural networks surpass the traditional restoration methods and become the mainstream in the computer vision area. To advance the performance of denoising algorithms, we propose a blind real image denoising network (SRMNet) by employing a hierarchical architecture improved from U-Net. Specifically, we use a selective kernel with residual block on the hierarchical structure called M-Net to enrich the multi-scale semantic information. Furthermore, our SRMNet has competitive performance results on two synthetic and two real-world noisy datasets in terms of quantitative metrics and visual quality. The source code and pretrained model are available at \url{https://github.com/TentativeGitHub/SRMNet}.
\end{abstract}
\section{Introduction}

Image denoising is a challenging ill-posed problem which also plays an important role in the pre-process of high-level vision task. In general, a corrupted image $Y$ could be represented as: 
\begin{equation}
Y = D(X) + n, \label{eq1}
\end{equation}
where $X$ is a clean image, $D(\cdot)$ denotes the degradation function and $n$ means the additive noise. Traditional model-based denoising methods, such as block-matching and 3D filtering (BM3D) \cite{01}, non-local means (NLM) \cite{02} are all based on the information of image priors. Although the conventional prior-based methods could handle most of denoising tasks and achieve acceptable performances, the key problems like computationally expensive and time-consuming hamper the efficiency of model-based methods. 

In recent years, learning-based methods \cite{04,16,17} surpass traditional prior-based methods in terms of inference time and denoising performance. The performance gain of learning-based methods especially CNN over the others is mainly attributed to their elaborately designed model or block. For example, residual learning \cite{03,04}, dense connection \cite{05,06}, residual dense block \cite{07,08}, attention mechanisms \cite{09,10,11}, channel attention block \cite{09,12}, and hierarchical architecture \cite{13,14,15}. However, these complex architectures cause the restoration models to waste more computation and the improvement is only a little. 

In this paper, we try to balance between the accuracy and computational efficiency of the model. First, we propose the hierarchical selective residual architecture which is based on the residual dense block with a more efficiency structure named selective residual block (SRB). Moreover, we use the multi-scale feature fusion with two different sampling methods (pixel shuffle \cite{18}, bilinear) based on the proposed M-Net to extract adequate and useful spatial feature information. For the reconstruction process, instead of using concatenation to fuse the feature maps with different resolutions, we adopt the selective kernel feature fusion (SKFF) \cite{19,20} to efficiently combine the features. Overall, the main contributions of this paper can be summarized as follows:

\begin{itemize}
  \item We propose the novel hierarchical architecture (M-Net) to denoise for both synthesized additive white Gaussian noise (AWGN) and real-world noise.
  \item We propose the efficient feature extraction block called selective residual block which is improved from the residual dense block for image super-resolution.
  \item We experiment on two synthetic image datasets and two real-world noisy datasets to demonstrate that our proposed model achieves the state-of-the-art in image denoising quantitatively and qualitatively even with less computational complexity.
\end{itemize}

\begin{figure*}[!t]
\centering
	\includegraphics[width=17cm]{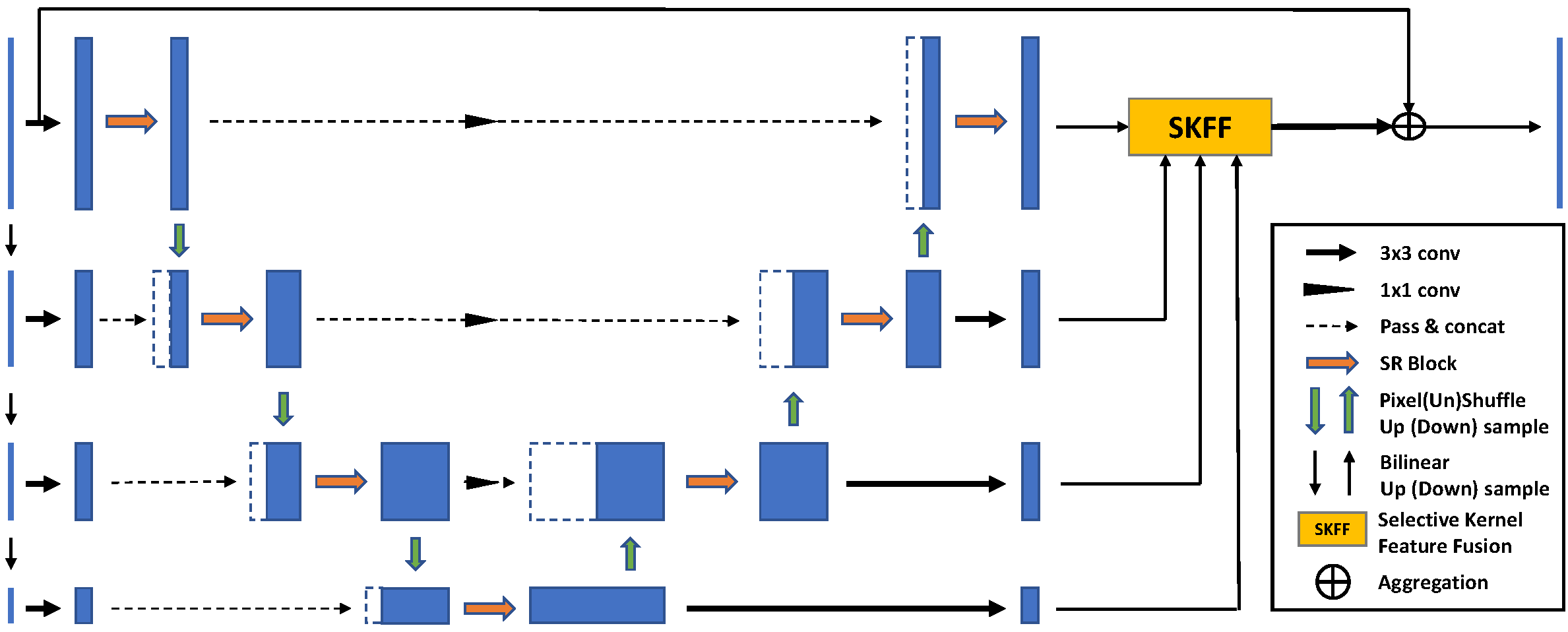}
	\caption{Proposed Selective Residual M-Net (SRMNet) architecture. The source code and component structure of the model could be found in the provided URL indicated in the abstract. We set the initial channel in each resolution to 96 after $3 \times 3$ convolution, and totally we have 4 layers in the proposed M-Net.}
	\label{SRMNet}
\end{figure*}

\section{Related Work}
In this section, we first are going to discuss the development of denoising. Then, we will describe the hierarchical architecture and the selective kernel which we apply in our proposed selective residual block.

\subsection{Image denoising} As aforementioned, traditional image denoising approaches are generally based on image priors or algorithms which are also called model-based methods, such as self-similarity \cite{01,02,21}, spare coding \cite{22,23} and dictionary learning \cite{22,25}. Currently, CNN-based denoisers have demonstrated state-of-the-art results \cite{03,04,13,14,15}. Moreover, the denoise models from \cite{04,11,14,15,16,17,44} only deal with signal-independent noise (e.g., AWGN, read noise), while the model from \cite{12,26,27,28,29} have the ability to process the real-world signal-dependent noise (e.g., shot noise, thermal noise).

\subsection{Hierarchical architecture} The hierarchical architecture is used to enrich the spatial information by extracting different sizes of features and make semantic information have more diversity. The most well-known hierarchical structure is the U-Net \cite{13} which has the great contribution from high-level to low-level vision tasks including image segmentation \cite{31,32}, image restoration \cite{14,15,30}, etc.

\subsection{Selective Kernel Network} Li \textit{et al.} proposed the selective kernel convolution that has two branches. One of the path utilizes normal naive $3 \times 3$ convolution kernel to extract features, and the other path adopts different kernel size (e.g. $5 \times 5$, $7 \times 7$) to obtain the larger receptive field. At the end of selective kernel convolution, they use softmax activation function to acquire the weights for two different features maps. Zamir \textit{et al.} \cite{20} were inspired by \cite{19} and applied it to multi-scale feature fusion for image enhancement tasks, which also achieve good results.

\section{Proposed Method}
In this section, we mainly introduce the proposed Selective Residual M-Net (SRMNet), and provide detailed explanations for each component of the model in the following subsections.

\subsection{M-Net}
The M-Net architecture is first proposed for medical image segmentation \cite{mehta2017m}. Adiga \textit{et al.} \cite{adiga2019fpd} use the same framework for fingerprint image denoising and also get good results. Compared with above two models, our proposed SRMNet has two improvements. 1) More diversity and plentiful multi-scale cascading features. The original M-Net used $2 \times 2$ max-pooling in both U-Net path and gatepost path, and then combined these two features together. Our SRMNet use pixel un-shuffle down-sampling in U-Net path and bilinear down-sampling for gatepost path, which makes the cascading features have more diversity. 2) Using different feature fusion methods to summarize the information in the decoder (reconstruction process). Actually, original M-Net has high-dimensional cascading features, especially the shallow layer in the model, which makes the M-Net have large number of parameters and high computational complexity. Therefore, they use some techniques such as batch normalization, reducing the size of input images and the dimension of input feature maps. In other words, the original M-Net is inappropriate to be directly applied to image denoising. To solve this problem, we use the selective kernel feature fusion (SKFF) method \cite{20} which does not concatenate each feature map but aggregates the weighting features.


\begin{figure*}[htbp] 
	\centering
	\includegraphics[width=11.0cm]{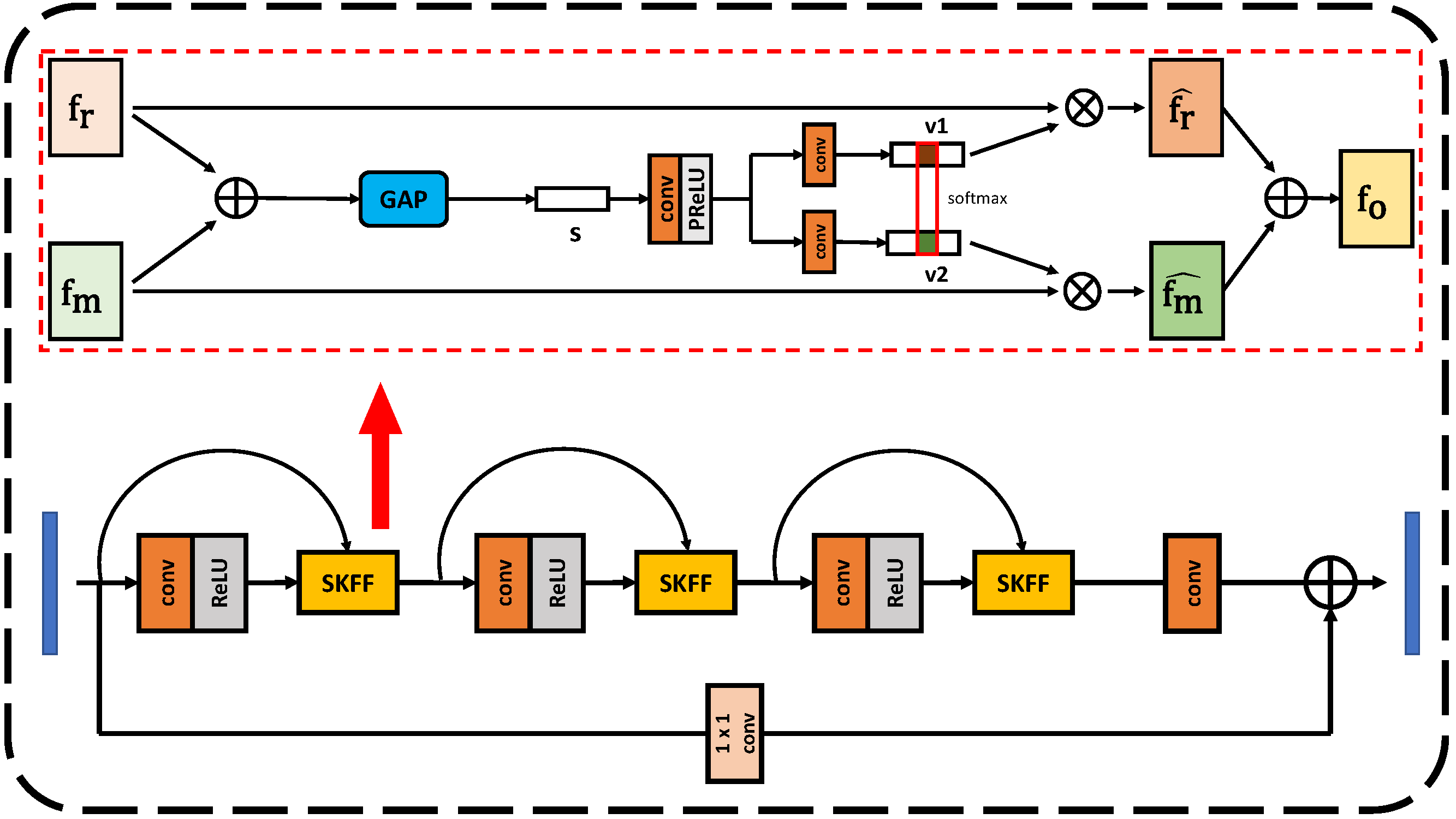}
  \caption{Illustration of the Selective Residual Block (SRB) in our SRMNet.}
  \label{SRB}
\end{figure*}

\subsection{SRMNet}
The proposed SRMNet for image denoising is shown in Fig. \ref{SRMNet}. We first use $3 \times 3$ weight sharing convolution in each resolution of corrupted input image acquired by doing the bilinear down-sampling from original-resolution input. Each layer has the proposed selective residual block (SRB) to extract high-level semantic information (more structure details will be illustrated in the next subsection). Then, we choose pixel unshuffled method as our down-sampling module at the end of SRB to obtain multi-scale feature maps. After that, the feature maps are concatenated with previous shallow features from bilinear down-sampling and keep going as normal U-Net process. The main purpose of choosing two different down-sample methods (pixel unshuffled and bilinear) is to make the cascaded features have more semantic information. Finally, using the SKFF (upper part of Fig. \ref{SRB}) to integrate features with different scales to reconstruct the denoised image. 

We optimize our SRMNet end-to-end with the Charbonnier loss \cite{33} for image denoising as follows:
\newcommand{\Lagr}{\mathcal{L}} 
\begin{equation}
\begin{aligned}
\Lagr_{char} = \sqrt{ ||\hat{X} - X||^2 + \varepsilon^2}, \label{eq2}
\end{aligned}
\end{equation}
where $\hat{X}, X\in\mathbb{R}^{B \times C \times H \times W}$ means the denoised and ground-truth images, respectively. $B$ is the batch size of training data, $C$ is the number of feature channels, $H$ and $W$ are the size of images. The constant $\varepsilon$ in Eq.(\ref{eq2}) are empirically set to $10^{-3}$.

\subsection{Selective Residual Block}
Fig. \ref{SRB} shows the architecture of the proposed SRB which is improved from the residual dense block (RDB) \cite{07}. In the framework of SRB, each residual block has two input features ($f_r, f_m \in \mathbb{R}^{C \times H \times W}$ in Fig. \ref{SRB}) which denote the residual feature and mainstream feature, respectively. These two features will do the SKFF by multiplying the corresponding feature descriptor vectors ($v_1, v_2 \in \mathbb{R}^{C \times 1 \times 1}$ which are generated from channel-wise statistics $s \in \mathbb{R}^{C \times 1 \times 1}$) to get the weighted features ($\hat{f}_r, \hat{f}_m \in \mathbb{R}^{C \times 1 \times 1}$). Finally, we aggregate two channel-weighted features $\hat{f}_r, \hat{f}_m$ together to acquire the output feature $f_o$ of single residual block. After a few residual blocks (e.g., 3), we use a $3 \times 3$ convolution and add the long skip connection with $1 \times 1$ convolution between the input and output.


\begin{figure}[htbp] 
	\centering
	\includegraphics[width=8.5cm]{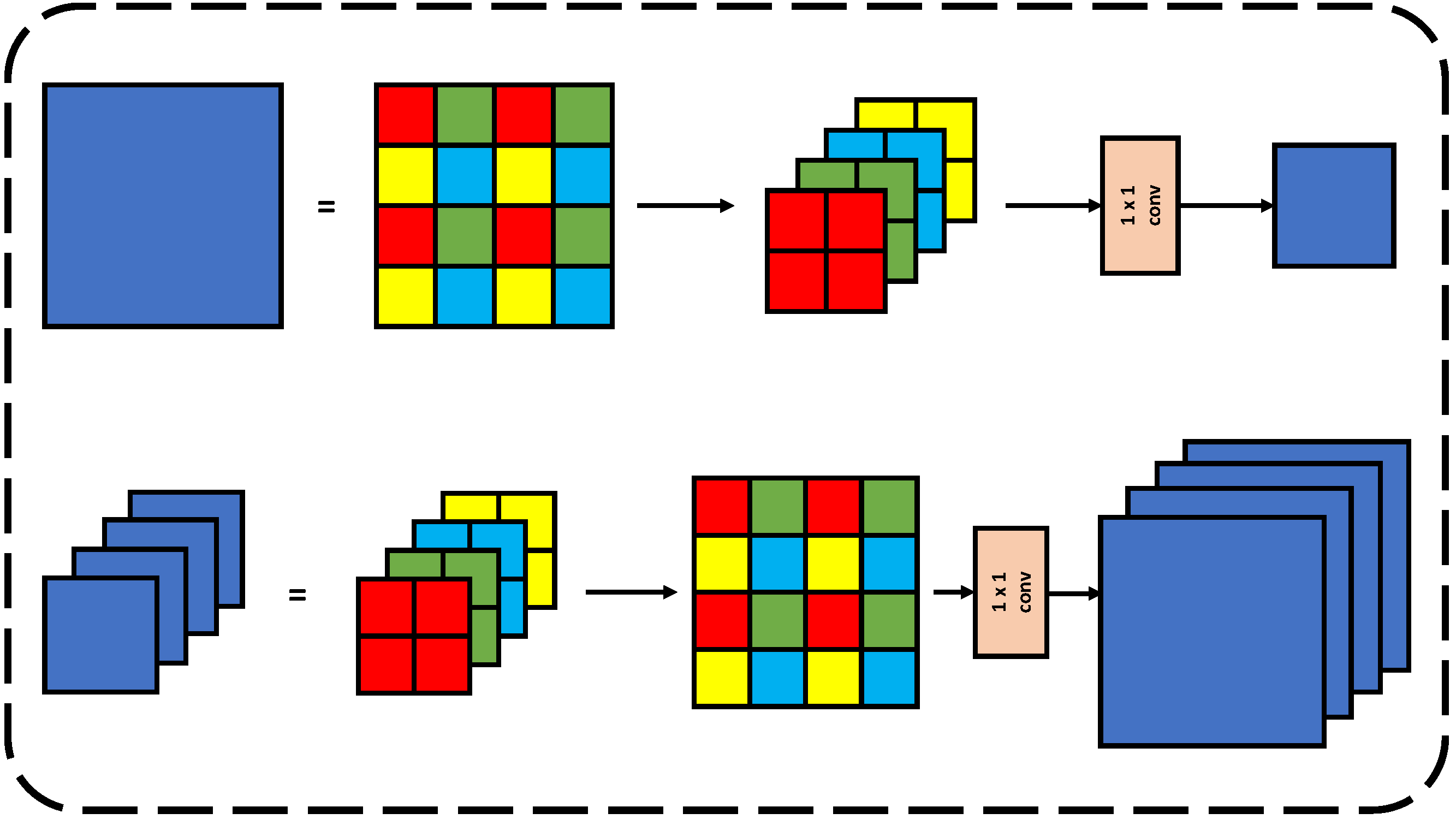}
  \caption{Resizing module with pixel (un)shuffle.}
  \label{PSM}
\end{figure}

\subsection{Resizing Module}
As for resizing module, we simply use bilinear down-sampling for input images $Y \in \mathbb{R}^{3 \times H \times W}$ (Y is the same as Eq. (\ref{eq1})), and use pixel unshuffled module shown in Fig. \ref{PSM} for shallow features $\in \mathbb{R}^{C \times H \times W}$after $3 \times 3$ convolution. Note that the size of the feature channel ($C$) before entering the resizing module is the same as output channel size but with smaller resolution (e.g. $W/2 \times H/2$). 

It should be noticed that the the feature map of lower layer contains both bilinear and pixel unshuffled features but the ingredients of bilinear features are obvious less than pixel unshuffled features. It may cause the unbalanced problem. In fact, the above unbalanced feature map will be passed through the SRB which could solve the unbalanced problem by increasing the weight of bilinear features. 

\section{Experiments}

\subsection{Experiment Setup}
\noindent\textbf{Implementation Details.} Our SRMNet is an end-to-end model and trained from scratch. The experiments conducted in this paper are implemented by PyTorch 1.8.0 with single NVIDIA GTX 1080Ti GPU.

\noindent\textbf{Evaluation Metrics.} For the quantitative comparisons, we consider the Peak Signal-to-Noise Ratio (PSNR) and Structure Similarity (SSIM) Index metrics.

\begin{table*}[!t] \footnotesize
\caption{\textbf{\underline{Gaussian color image denoising}}. Image denoising results on CBSD68 dataset \cite{35} and Kodak24 dataset \cite{36}. The best and second best scores are \textbf{highlighted} and \underline{underlined}, respectively. All of scores are the average values of the whole dataset. The last column shows floating-point operations per second (FLOPs) which is conducted on $256\times256$ color images.}
\label{denoise_table}
\begin{center}
\begin{tabular}{l  || cc | cc| cc || cc | cc| cc || c}
\toprule[1.5 pt]
\multirow{3}{*}{\textbf{Methods}}&\multicolumn{6}{c||}{CBSD68 \cite{35}}&\multicolumn{6}{c||}{Kodak24 \cite{36}}&\multirow{3}{*}{FLOPs}\\
&\multicolumn{2}{c}{$\sigma = 10$}&\multicolumn{2}{c}{$\sigma = 30$}&\multicolumn{2}{c||}{$\sigma = 50$}&\multicolumn{2}{c}{$\sigma = 10$}&\multicolumn{2}{c}{$\sigma = 30$}&\multicolumn{2}{c||}{$\sigma = 50$}\\
&\multicolumn{1}{c}{PSNR}&\multicolumn{1}{c}{SSIM}&\multicolumn{1}{c}{PSNR}&\multicolumn{1}{c}{SSIM}&\multicolumn{1}{c}{PSNR}&\multicolumn{1}{c||}{SSIM}&\multicolumn{1}{c}{PSNR}&\multicolumn{1}{c}{SSIM}&\multicolumn{1}{c}{PSNR}&\multicolumn{1}{c}{SSIM}&\multicolumn{1}{c}{PSNR}&\multicolumn{1}{c||}{SSIM}\\
\midrule[1.5 pt]
BM3D \cite{01}
&35.89&0.951&29.71&0.843&27.36&0.763&33.32&0.956&27.75&0.773&25.60&0.686&-\\
IrCNN \cite{17}
&36.06&0.953&30.22&0.861&27.86&0.789&36.70&0.945&31.24&0.858&28.92&0.794&27G\\
FFDNet \cite{16}
&36.14&0.954&30.31&0.860&27.96&0.788&36.80&0.946&31.39&0.860&29.10&0.795&18G\\	
DnCNN \cite{04}
&36.12&0.951&30.32&0.861&27.92&0.788&36.58&0.945&31.28&0.858&28.94&0.792&36G\\
DHDN \cite{14}
&36.05&0.953&30.12&0.858&27.71&0.787&37.30&0.951&\textbf{31.98}&0.874&29.72&0.817&1019G\\
RNAN \cite{11}
&36.43&-&30.63&-&26.83&-&37.24&-&31.86&-&29.58&-&-\\
DIDN \cite{44}
&\underline{36.48}&\underline{0.957}&30.71&0.870&28.35&0.804&\textbf{37.32}&0.950&31.97&0.872&29.72&0.816&1121G\\
RDN \cite{07}&36.47&-&30.67&-&28.31&-&\underline{37.31}&-&31.94&-&29.66&-&1490G\\
RDUNet \cite{15}
&\textbf{36.48}&0.951&\underline{30.72}&\underline{0.872}&\underline{28.38}&\underline{0.807}&37.29&\underline{0.951}&31.97&\underline{0.874}&\underline{29.72}&\underline{0.818}&807G\\
\midrule[1.5 pt]
\textbf{SRMNet (Ours)}
&36.46&\textbf{0.961}&\textbf{30.72}&\textbf{0.878}&\textbf{28.38}&\textbf{0.814}&37.29&\textbf{0.957}&\underline{31.97}&\textbf{0.882}&\textbf{29.72}&\textbf{0.826}&285G\\
\bottomrule[1.5pt]
\end{tabular}
\end{center}
\end{table*}

\subsection{Experiment Datasets}
\noindent\textbf{Gaussian Color Image Denoising.} For Gaussian denoising, we use the same experimental setup as image denoising \cite{14,15} and train our model on image super-resolution DIV2K \cite{39} dataset which has 800 and 100 high-quality (the average resolution is about $1920 \times 1080$) images for training and validation, respectively. We randomly crop 100 patches with size $256 \times 256$ for each training image and randomly add AWGN to the patches with noise level from $\sigma$ = 5 to 50. Evaluation is conducted on noise levels 10, 30, 50 on CBSD68 \cite{35} and Kodak24 \cite{36}. It should be noted that our model does not know the noise level in the testing, which means SRMNet is the blind denoising model.

\noindent\textbf{Real-World Image Denoising.} To train our SRMNet for real-world denoising, we follow \cite{20, 30} to use 320 high-resolution images of SIDD dataset \cite{40}. Evaluations also follow aforementioned methods, which is to perform the test on 1280 validation patches from the SIDD dataset \cite{40} and 1000 patches from the DND benchmark dataset \cite{41}. The resolution of all patches is $256 \times 256$ in both training and testing.

\begin{table}[!t]\footnotesize
\caption{\textbf{\underline{Real-world image denoising}}. Image denoising result on SIDD \cite{40} and DND \cite{41} datasets. $^*$ denotes the method used additional training data. The proposed SRMNet is only trained on the SIDD images and then tested on DND.}
\label{real_world_table}
\begin{center}
\setlength{\tabcolsep}{3.5mm}{
\begin{tabular}{l || cc || cc}
\toprule[1.5 pt]
\multirow{2}{*}{\textbf{Methods}}&\multicolumn{2}{c||}{SIDD \cite{40}}&\multicolumn{2}{c}{DND \cite{41}}\\
&PSNR&SSIM	&PSNR&SSIM\\
\midrule[1.5 pt]
DnCNN \cite{04}&23.66&0.583&32.43&0.790\\
BM3D \cite{01}&25.65&0.685&34.51&0.851\\
CBDNet$^*$ \cite{26}&30.78&0.801&38.06&0.942\\
RIDNet$^*$ \cite{27}&38.71&0.951&39.26&0.953\\
DAGL \cite{46}&38.94&0.953&39.77&0.956\\
AINDNet$^*$ \cite{45}&38.95&0.952&39.37&0.951\\
VDN \cite{28}&39.28&0.956&39.39&0.952\\
DeamNet$^*$ \cite{43}&39.35&0.955&39.63&0.953\\
SADNet$^*$ \cite{29}&39.46&0.957&39.59&0.952\\
CycleISP$^*$ \cite{12}&39.52&0.957&39.56&\underline{0.956}\\
DANet+$^*$ \cite{42}&39.47&0.957&39.58&0.955\\
MPRNet \cite{30}&39.71&0.958&\underline{39.80}&0.954\\
MIRNet \cite{20}&\textbf{39.72}&\underline{0.959}&\textbf{39.88}&\textbf{0.956}\\
\midrule[1.5 pt]
\textbf{SRMNet (Ours)}&\underline{39.72}&\textbf{0.959}&39.44&0.951\\
\bottomrule[1.5pt]
\end{tabular}}
\end{center}
\end{table}

\subsection{Image Denoising Performance}
\begin{figure*}[!t]
\centering
	\subfigure{
	\begin{minipage}{2.1cm}
	\includegraphics[width=2cm]{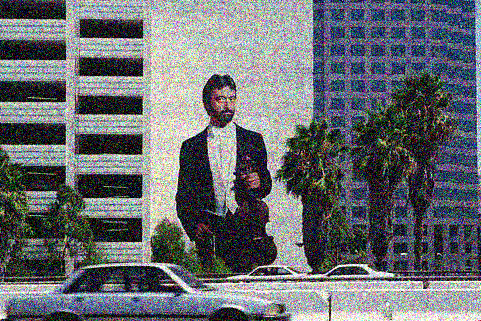}
	\vspace*{-3mm}
	\caption*{\small{15.26/0.414}}
	\vspace*{-4mm}
	\caption*{\small{Noisy Image}}
	\end{minipage}
	\begin{minipage}{2.1cm}
	\includegraphics[width=2cm]{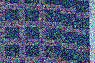}
	\vspace*{-3mm}
	\caption*{\small{15.26/0.414}}
	\vspace*{-4mm}
	\caption*{\small{Noisy Patch}}
	\end{minipage}
	\begin{minipage}{2.1cm}
	\includegraphics[width=2cm]{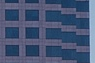}
	\vspace*{-3mm}
	\caption*{\small{PSNR/SSIM}}
	\vspace*{-4mm}
	\caption*{\small{Reference}}
	\end{minipage}
	\begin{minipage}{2.1cm}
	\includegraphics[width=2cm]{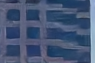}
	\vspace*{-3mm}
	\caption*{\small{27.59/0.840}}
	\vspace*{-4mm}
	\caption*{\small{DnCNN \cite{04}}}
	\end{minipage}
	\begin{minipage}{2.1cm}
	\includegraphics[width=2cm]{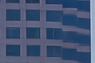}
	\vspace*{-3mm}
	\caption*{\small{28.18/0.850}}
	\vspace*{-4mm}
	\caption*{\small{DHDN \cite{14}}}
	\end{minipage}
	\begin{minipage}{2.1cm}
	\includegraphics[width=2cm]{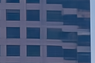}
	\vspace*{-3mm}
	\caption*{\small{28.25/0.858}}
	\vspace*{-4mm}
	\caption*{\small{DIDN \cite{44}}}
	\end{minipage}
	\begin{minipage}{2.1cm}
	\includegraphics[width=2cm]{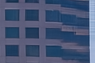}
	\vspace*{-3mm}
	\caption*{\small{\underline{28.31}/\underline{0.866}}}
	\vspace*{-4mm}
	\caption*{\small{RDUNet \cite{15}}}
	\end{minipage}
	\begin{minipage}{2.1cm}
	\includegraphics[width=2cm]{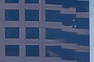}
	\vspace*{-3mm}
	\caption*{\small{\textbf{28.39}/\textbf{0.874}}}
	\vspace*{-4mm}
	\caption*{\footnotesize{\textbf{SRMNet (Ours)}}}
	\end{minipage}
	}%
	\vspace*{-2mm}
	\subfigure{
	\begin{minipage}{2.1cm}
	\includegraphics[width=2cm]{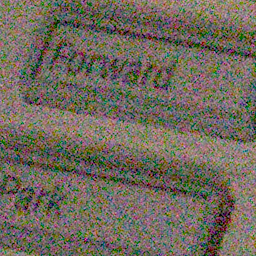}
	\vspace*{-3mm}
	\caption*{\small{18.25/0.239}}
	\vspace*{-4mm}
	\caption*{\small{Noisy}}
	\end{minipage}
	\begin{minipage}{2.1cm}
	\includegraphics[width=2cm]{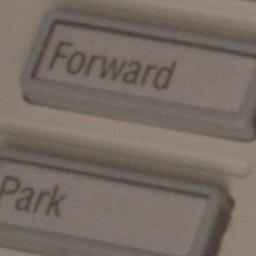}
	\vspace*{-3mm}
	\caption*{\small{PSNR/SSIM}}
	\vspace*{-4mm}
	\caption*{\small{Reference}}
	\end{minipage}
	\begin{minipage}{2.1cm}
	\includegraphics[width=2cm]{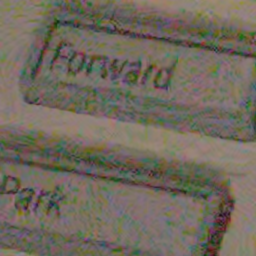}
	\vspace*{-3mm}
	\caption*{\small{25.75/0.833}}
	\vspace*{-4mm}
	\caption*{\small{BM3D \cite{01}}}
	\end{minipage}
	\begin{minipage}{2.1cm}
	\includegraphics[width=2cm]{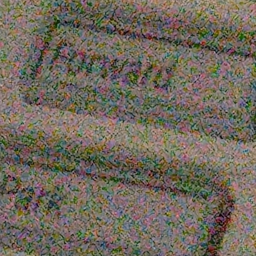}
	\vspace*{-3mm}
	\caption*{\small{20.76/0.261}}
	\vspace*{-4mm}
	\caption*{\small{DnCNN \cite{04}}}
	\end{minipage}
	\begin{minipage}{2.1cm}
	\includegraphics[width=2cm]{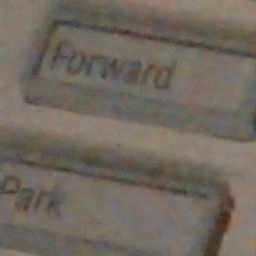}
	\vspace*{-3mm}
	\caption*{\small{28.84/0.0.858}}
	\vspace*{-4mm}
	\caption*{\small{CBDNet \cite{26}}}
	\end{minipage}
	\begin{minipage}{2.1cm}
	\includegraphics[width=2cm]{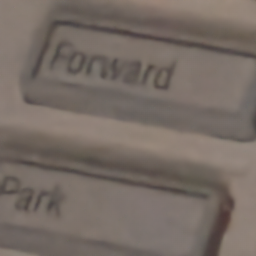}
	\vspace*{-3mm}
	\caption*{\small{35.57/0.934}}
	\vspace*{-4mm}
	\caption*{\small{RIDNet \cite{27}}}
	\end{minipage}
	\begin{minipage}{2.1cm}
	\includegraphics[width=2cm]{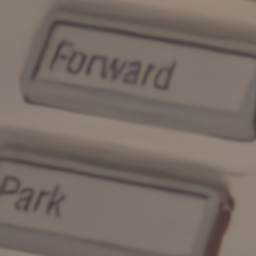}
	\vspace*{-3mm}
	\caption*{\small{\underline{36.75}/\underline{0.950}}}
	\vspace*{-4mm}
	\caption*{\small{CycleISP \cite{12}}}
	\end{minipage}
	\begin{minipage}{2.1cm}
	\includegraphics[width=2cm]{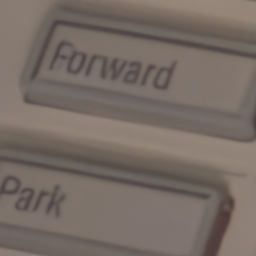}
	\vspace*{-3mm}
	\caption*{\small{\textbf{36.86}/\textbf{0.952}}}
	\vspace*{-4mm}
	\caption*{\footnotesize{\textbf{SRMNet (Ours)}}}
	\end{minipage}
	}%
\caption{Visual comparisons of image denoising on the CBSD68 \cite{35} (upper row) and SIDD \cite{40} (bottom row) datasets for color Gaussian and real image denoising, respectively. Due to the page limits, more visual results for different datasets could be found in our github page.}
\label{denoise_images}
\end{figure*}

\noindent\textbf{Gaussian Color Image Denoising.} In Table~\ref{denoise_table} and Fig.~\ref{denoise_images}, we compare our SRMNet with the prior-based method (e.g., BM3D \cite{01}), CNN-based methods (e.g., DnCNN \cite{04}, IrCNN \cite{17}, FFDNet \cite{16}) and the models which are based on RDB (e.g., DHDN \cite{14}, DIDN \cite{44}, RDN \cite{07}, RDUNet \cite{15}). According to Table~\ref{denoise_table}, we could observe three things: 1) The proposed SRMNet achieves state-of-the-art quantitative scores, especially for the difficult Gaussian noise levels (e.g., 30, 50). 2) Compared to RDB-based methods (DHDN, DIDN, RDN, RDUNet), our SRMNet has the least FLOPs ($\downarrow 20\%$) among the five models, and still keeps the best scores because of the efficient SRB design. 3) The SSIM scores of the SRMNet are the best in both CBSD68 and Kodak24 datasets for each noise level, which means that our denoised images are more perceptually faithful. We think it is attributed to the M-Net design which could gain more spatial details in the training process.

\noindent\textbf{Real-World Image Denoising.} For real image denoising, we evaluate the performance of 13 image denoising approaches on real-world noise datasets (SIDD and DND) in Table~\ref{real_world_table}. Compared to the previous state-of-the-art CNN-based method \cite{20} in SIDD dataset, our model gains the same scores with MIRNet but less computational complexity (e.g., FLOPs) and time cost. More specifically, Table~\ref{compare_table} shows our SRMNet only use 36.3\% of MIRNet's FLOPs and about three times faster than MIRNet. Fig.~\ref{denoise_images} also displays the visual results for real image denoising on SIDD. Our SRMNet effectively removes noise and the denoised images are visually closer to the ground-truth.

\begin{table}[!t]\small
\caption{Comparison of the PSNR and FLOPs for MIRNet \cite{20}, MPRNet \cite{30} and our method (SRMNet). We did the test on SIDD validation set which has 1,280 patches where FLOPs are estimated on the input with shape of $1 \times 3 \times 256 \times 256$. The inference times are measured on the computer equipped with NVIDIA GTX 1080Ti GPU.}
\label{compare_table}
\begin{center}
\setlength{\tabcolsep}{1mm}{
\begin{tabular}{l || c || rr || rr}
\toprule[1.5 pt]
{\textbf{Methods}}&\multicolumn{1}{c||}{PSNR}&\multicolumn{2}{c||}{FLOPs (G)}&\multicolumn{1}{r}{Time (ms)}&\multicolumn{1}{r}{Speedup}\\
\midrule[1.5 pt]
MIRNet \cite{20}&39.72&787.04&100\%&212.94&$1\times$\\
MPRNet \cite{30}&39.71&573.88&72.9\%&128.77&$1.65\times$\\
SRMNet (Ours)&39.72&285.36&36.3\%&71.29&$2.98\times$\\
\bottomrule[1.5pt]
\end{tabular}}
\end{center}
\end{table}

\section{Conclusion}
In this paper, we present the SRMNet architecture and achieve state-of-the-art performances on image denoising. The M-Net design has the advantage of enriching features with different resolutions by concatenating the results after pixel unshuffle and bilinear down-sampling. Moreover, we proposed the SRB, which is an efficient block compared with the RDB. Our future works are going to focus on different restoration tasks such as image deblurring and image deraining.

\bibliographystyle{IEEEbib}
\fontsize{8.9pt}{8.9pt}
\selectfont
\bibliography{references}

\end{document}